# Surface-enhanced Raman spectroscopy in 3D electrospun nanofiber mats coated with gold nanorods

Andrea Camposeo,[1,2,*] Donatella Spadaro,[3] Davide Magrì,[2] Maria Moffa,[1] Pietro G. Gucciardi,[3] Luana Persano,[1,2,*] Onofrio M. Maragò,[3,*] Dario Pisignano[1,4]

[1]*Istituto Nanoscienze-CNR, Euromediterranean Center for Nanomaterial Modelling and Technology (ECMT), via Arnesano, I-73100, Lecce, Italy.*

[2]*Soft Materials and Technologies SRL, via Arnesano 16, I-73100, Lecce, Italy.*

[3]*CNR-IPCF, Istituto per i Processi Chimico-Fisici, I-98158, Messina, Italy.*

[4]*Dipartimento di Matematica e Fisica "Ennio De Giorgi", Università del Salento, via Arnesano, I-73100, Lecce, Italy.*

*Corresponding authors: A.C.: andrea.camposeo@nano.cnr.it; O. M. M.: marago@ipcf.cnr.it; L.P.: luana.persano@nano.cnr.it

**Abstract**

Nanofibers functionalized by metal nanostructures and particles are exploited as effective flexible substrates for SERS analysis. Their complex three-dimensional structure may provide Raman signals enhanced by orders of magnitude compared to untextured surfaces. Understanding the origin of such improved performances is therefore very important for pushing nanofiber-based analytical technologies to their upper limit. Here we report on polymer nanofiber mats which can be exploited as substrates for enhancing the Raman spectra of adsorbed probe molecules. The increased surface area and the scattering of light in the nanofibrous system are individually analyzed as mechanisms to enhance Raman scattering. The deposition of gold nanorods on the fibers further amplifies Raman signals due to SERS. This study suggests that Raman signals can be finely tuned in intensity and effectively enhanced in nanofiber mats and arrays by properly tailoring the architecture, composition, and light-scattering properties of the complex networks of filaments.

## Introduction

The recent advances in nanophotonics, unveiling the mechanisms at the base of the interaction of light with nanostructured materials, have established novel paradigms for analytical and bio-analytical applications [1-3]. In these fields, the use of light is highly desirable, being noninvasive and providing fast response, high sensitivity and low power consumption. In particular, the efficiency and sensitivity of photonics-based analytical methods can be significantly enhanced by controlling the strength of light-analyte coupling [4, 5]. Interesting optical architectures to this aim include optical micro- and nanocavities [6-8], photonic crystals [9, 10], sub-wavelength waveguides and optical nanofibers [11-13], metallic nanoparticles and nanostructures [14-16], and hybrid systems [17, 18]. Embedding these structures in analytic devices can significantly enhance light absorption, scattering and fluorescence, and enable targeted detection at extremely low concentrations and with sensitivities approaching the single molecule level [14, 19].

In this framework, Raman spectroscopy is a highly efficient tool, which allows for the identification of a variety of materials [20]. This technique relies on the inelastic scattering of light by molecules or materials, yielding information on their vibrational modes, structural and electronic properties. It is fast and non-destructive, offering high sensitivity in the recognition of different species [21], and it can be integrated with advanced optical microscopy, optical fibers, miniaturized lasers and other photonic devices to improve diagnostic performances [22-24]. For instance, metal nanoparticles have been used to amplify Raman signals by several orders of magnitude through their plasmonic response [25]. Surface-enhanced Raman scattering (SERS) spectroscopy relies on finding that Raman signals from molecules adsorbed on rough surfaces or nanoparticles made of noble metal (typically Au and Ag) are significantly enhanced upon laser illumination [5, 26-29]. This enhancement is due to two phenomena, i.e. a weak chemical and a stronger electromagnetic mechanism [30-32], related to the resonant interaction of light with localized surface plasmons (LSPs) excited in the nanoparticles. In fact, the aggregation of metal nanoparticles creates so-called

'hot spots', namely nanocavity regions between adjacent particles that further amplify the Raman signal of adsorbed molecules [33, 34]. In recent years, SERS has been successfully applied in surface chemistry and electrochemistry [35, 36], forensics [37], chemical sensors [36, 38], and biosensors [36-37, 39-40].

Au and Ag plasmonic substrates with well-controlled architecture and providing SERS enhancement factors exceeding $10^{10}$ have been reported [41-43]. Most of such substrates are realized by complex and multi-step lithographic and nanofabrication procedures, and on rigid and planar surfaces, which can severely limit the application of these technologies in point-of-care diagnostic applications [44]. In this respect, SERS substrates should be preferably flexible and cheap, robust enough to be used with a variety of fluids (air, gas or liquids), and compatible with lab-on-chip integration. Recently, polymer nanofibers have emerged as effective substrates for SERS [45, 46], allowing Raman signals to be amplified by many orders of magnitude. Polymer nanofibers exhibit a very high surface-to-volume ratio, and they can be produced with tailored optical properties by electrospinning [47, 48] and functionalized with a variety of metal nanoparticles. Electrospun mats can be produced in continuous runs, with size ranging from few cm$^2$ to the scale of m$^2$, by using either single or multiple electrified jets and needleless systems [49, 50]. Moreover, electrospun nanofibers can be made by biocompatible and environmentally friendly materials, and can be arranged in complex architectures, thus forming suitable substrates for tissue engineering and air and liquid filtration [51]. Therefore, the development of SERS systems based on electrospun nanofibers would potentially allow sensing capability to be integrated in scaffolds and filter devices.

Various substrates for SERS have been based on the embodiment of Au and Ag nanoparticles in fibers following dispersion in electrospun solutions [45, 52] or *in-situ* synthesis [53]. In other approaches, the fibers have been grafted or decorated by metal nanoparticles or nanosheets [54-56]. Demonstrated applications include the detection of organic pollutant [57], of biogenic amines and bacteria [58, 59], and the use of fibers in microfluidic devices [60]. However, the origin of the enhancement of the SERS signals by electrospun fibers is still not well understood. The comparison with flat substrates having similar composition suggests a role played by the three-dimensional (3D) structure of fiber mats and by the aggregation and assembly of the used nanoparticles [45, 54, 56]. However, embedding metal particles within polymer matrices as in many previous studies does not allow the contribution of the fiber structure to be untangled.

Here, we report on the investigation of the Raman and SERS spectra enhancement in polymer nanofibers following physical deposition of gold nanorods (GNRs). 3D substrates made of poly(methylmethacrylate) (PMMA) with average diameters of about 400 nm were produced by electrospinning and functionalized by a prototype molecular analyte. Interestingly, the Raman signal is enhanced by an order of magnitude by fibers with respect to films. GNR deposition at very low density provides an additional signal gain by a factor 10, due to SERS effect. These results are rationalized accounting for the 3D structure of the fiber mat and for the scattering of the incident light by the fibers, establishing new design rules for improving the performances of 3D nano- and micro-architectures for Raman and SERS applications.

## Experimental Details

*Electrospinning*. Fibers were spun from a solution of PMMA (M.W. 120,000 g/mol, Sigma Aldrich) in formic acid (Alfa Aesar) with concentration $2.6\text{-}2.9 \times 10^{-5}$ M. The solution was stirred for 24 h, and loaded in a 1 mL plastic syringe tipped by a 27G stainless steel needle. A positive bias of +9 kV was applied to the spinneret by a high voltage power supply (Glassman High Voltage). Glass coverslips ($18 \times 18$ mm$^2$, thickness 150 μm) were placed on a metallic plate ($10 \times 10$ cm$^2$), used as collector and biased at -6 kV (distance needle-collector 20 cm), while the solution was delivered at constant rate (15 μL/min) by a microprocessor dual drive syringe pump (Harvard Apparatus). Electrospinning was carried out in ambient conditions (temperature 21 °C and relative humidity about 30%). Reference films were realized by spin-casting the PMMA solution on glass coverslips at 4000 rpm.

*Nanofiber characterization*. The morphology of nanofibers was investigated by optical, electron and atomic force microscopy (AFM). The former was carried out by using an up-right optical microscope (BX51, Olympus) in dark-field mode. Scanning electron microscopy (SEM) was performed by using a NOVA NANOSEM 450 system (FEI), with an acceleration voltage of 5 kV. AFM experiments were performed by a Multimode head equipped with a Nanoscope IIIa electronic controller (Veeco). The topography of arrays of fibers was imaged in tapping mode, using Phosphorous-doped Si cantilevers with resonance frequency of 377 kHz.

*Raman/SERS spectroscopy*. Raman/SERS spectroscopy measurements were carried out with a Jobin-Yvon HR800 micro-spectrometer coupled to a linearly polarized He-Ne laser emitting at 632.8 nm (6.7 mW excitation power). Light was focused on a spot of ~1.5 μm diameter by a 50× long-working-distance objective (numerical aperture = 0.5). This objective was also used to collect the back-scattered radiation. The depth of focus of the excitation and collection optical system was estimated to be about 3 μm at the operational wavelength (632.8 nm). Spectral analysis was accomplished by using a 600 lines/mm grating. A Silicon Charge Coupled Device (CCD) camera was used for light detection with typical integration times ranging from 10 to 30 s.

As SERS substrates, we used GNRs with size 25×75 nm (diameter×length), purchased from Nanopartz. The GNRs were dispersed in deionized (DI) water at a pH=3-5 with a concentration of about $8 \times 10^{10}$ GNRs $mL^{-1}$, and coated with hexadecyltrimethylammonium bromide (CTAB) surfactant to prevent spontaneous re-aggregation. CTAB is a surfactant widely used for the synthesis and dispersion of GNRs [61]. It provides good control over the growth and morphology of nanoparticles and stabilizes them against self-aggregation, through the formation of a chemisorbed bilayer on their surface [62]. While this is crucial during growth and stabilization in solution, it might be detrimental for SERS applications as CTAB can shield the GNR surface from analyte molecules. A promising way to improve SERS performance of CTAB capped GNRs is to exploit ligand exchange, where CTAB destabilization is obtained by taking advantage of sulfur affinity for gold and the CTAB bilayer might be displaced by the desired analyte molecules [62]. Moreover, the use of CTAB as capping layer of GNRs has been reported to have beneficial effects for both adsorption of GNRs on polymer filaments and SERS signal enhancement [46]. The presence of CTAB is expected to provide local hydrophylicity, thus favoring the adsorption of analytes nearby the GNR hot-spots.

The optical extinction spectrum of the used GNRs is shown in Figure S1 of the Electronic Supplementary Material (ESM). The LSP resonance is peaked at about 700 nm. GNRs were drop cast by delivering a 10 µL dispersion volume onto a surface of 250 $mm^2$. Methylene blue (MB, $C_{16}H_{18}N_3SCl$) was used as probe molecule in Raman/SERS experiments. MB is a heterocyclic aromatic compound that is often employed in SERS studies [28, 29]. The most characteristic vibrational modes of MB are in the 400-1650 $cm^{-1}$ range [29]. As typical for dyes, the electronic resonances of MB favor spectra of molecules adsorbed on substrates with low or no plasmonic activity to be collected, by exploiting the enhancement due to resonant Raman scattering effects [30]. In our experiments a 10 µL drop of MB aqueous solution at different concentrations ($10^{-8}$ M- $10^{-4}$ M) was cast on the fiber mats. To evaluate the Raman signal gain provided by the nanofibers alone and the SERS signal gain provided by the GNR-coated fibers, a PMMA spin-cast film and a glass slide were used as flat substrates for either Raman (without GNRs) or SERS (with GNRs) control experiments.

## Results and Discussion

### Nanofiber morphology

Figure 1 shows the morphology of the obtained PMMA fibers. The corresponding distribution of fiber diameters is displayed in the inset of Figure 1a, featuring an average value of 400 nm and a full width at half maximum of 300 nm. Various phenomenological and analytical models have been developed to predict electrospun fiber properties [63-65]. Here, thicker fibers were achieved by increasing even slightly the polymer solution concentration, whereas thinner fibers were obtained by the addition of an organic salt, tetrabutylammonium iodide (TBAI, Figure S2 of the ESM). Indeed, the addition of organic salts increases the solution conductivity and decreases the surface tension, thus favoring the formation of more stable and continuous electrospun jets [66].

The morphology of the deposited mats is displayed in Figure 1b, clearly showing the complex network formed by the fibers which constitute a truly 3D substrate (the corresponding morphology of a reference spin cast film in shown in Fig. 1c). The available surface of thick nanofiber mats can be orders of magnitude higher than that of a flat substrate, and it can be deployed for the development of highly sensitive optical sensors [67]. In this work we focus on mats composed by a few layers of fibers (typically less than 5 layers), that are intentionally chosen to study the effect of the complex network on the Raman spectra of probe molecules.

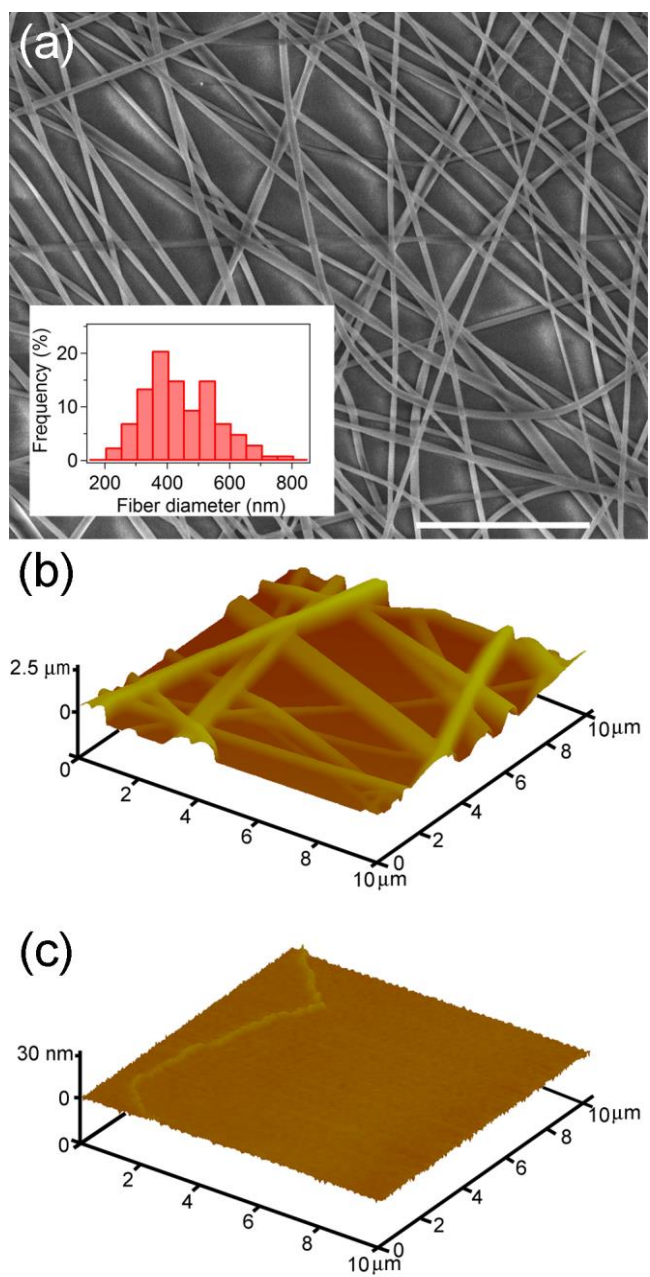

**Figure 1.** (a) SEM micrograph of a mat of electrospun PMMA fibers. Polymer solution concentration = $2.6 \times 10^{-5}$ M. Scale bar: 20 μm. Inset: fiber diameter distribution. (b)-(c) AFM topographic maps of a fiber mat (b) and of a spin-cast film (c), respectively.

## Raman spectra on nanofibers

Figure 2a and 2b show typical Raman spectra of MB molecules collected from electrospun nanofibers and spin-cast films, respectively. The spectra display several active modes, attributed to both the PMMA substrate and MB (see Table 1 and Figure S3 of ESM) [28, 68, 69]. Comparing results obtained by the two different substrates indicates that the surface topography greatly impacts on the intensity of the Raman spectra. In particular, taking the peak at 1618 cm$^{-1}$ as reference, an increase by about 15 times is observed in the Raman signal.

Two main effects can contribute to such enhancement. The first mechanism is related to the 3D geometry of the fibers, which enhances the surface available to molecular adsorption and to the consequent generation of detectable Raman signal.

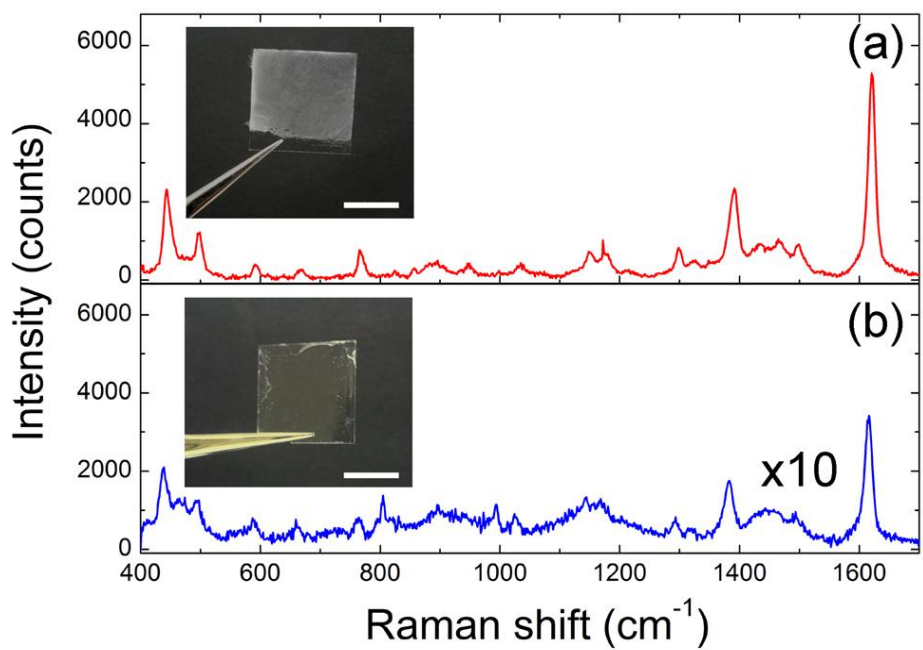

**Figure 2.** Raman spectra of MB cast from an aqueous solution on a layer of PMMA nanofibers (a) and on a spin-coated PMMA film (b). The Raman spectrum displayed in panel (b) is multiplied by a factor 10. The spectra are collected by an integration time of 30 s and the fluorescence background has been subtracted. Insets: corresponding sample photographs. Scale bars: 1 cm.

**Table 1**: Assignment of selected vibrational frequencies measured in Raman and SERS experiments.

| Wavenumber (cm$^{-1}$) | Molecular system | Mode assignment | Reference |
|---|---|---|---|
| 1618 | MB | ν(C-C)ring | [28] |
| 1456 | PMMA | δ(C-H) | [68] |
| 1441 | MB | ν$_{asym}$(C-N) | [28] |
| 1180 | MB | ν(C-N) | [28] |
| 1067 | MB | β(C-H) | [28] |
| 970 | PMMA | α-CH$_3$ rock | [68] |
| 813 | PMMA | ν(CH$_2$) | [69] |
| 604 | PMMA | C-O stretch | [68] |
| 497, 445 | MB | δ(C-N-C) | [28] |

At single-fiber level, this geometrical effect provides an area enhanced by ~3 times, given by the ratio of the surface ($A_{lat}$) of a nanofiber of radius $r$ and length $l$, $A_{lat} = 2\pi r l$, and its 2D projection, $A_{2D} = 2rl$. This ratio can be roughly multiplied by a further factor $n$ if we illuminate sample regions in which $n$ fibers overlap. From the AFM and SEM images the average number of fibers put on top of other fibers turns out to be $\bar{n} = 3$. The expected average amplification of the Raman signal due to the enhanced area is therefore of the order of $G_{geo} = \bar{n} \, A_{lat}/A_{2D} \sim 10$. This calculation provides an upper value for $G_{geo}$, since it does not account for the porosity of the fiber mats. A more accurate estimate is obtained by using the AFM topographic maps and by calculating the lateral surface of different areas of the nanofiber mats, having size comparable to the excitation laser spot size. Such analysis provides an average value of $G_{geo}$ around 7.

The second effect is related to the multiple scattering of incident light by the nanofibers. While in a flat substrate the incident light interacts with the layer of the probe molecule basically through a single pass, in fibrous materials photons can be repeatedly scattered, thus enhancing the probability to give rise to a Raman scattering process by the molecules adsorbed on the surface of the fibers, as schematized in Figure 3a,b. Indeed, observing fibers by dark-field optical microscopy in a configuration similar to that of Raman measurements (inset of Figure 3c), clearly highlights that a significant amount of light deviates from its incidence direction, even being possibly back-scattered. This effect is better illustrated in Figure 3c, where the light scattering form factor, being related to the intensity of the scattered light, is calculated for a fiber with 400 nm diameter in the Rayleigh-Gans approximation [70] as a function of the angle, $\theta$, in a plane perpendicular to the fiber longitudinal axis (scheme in Fig. 3c). A non-negligible amount of the incident light is diffused at large angles (>90°), where interactions with adjacent nanofibers are more likely. Multiple scattering processes are also expected to increase the light back-scattered within the numerical aperture of the collecting objective. Although a quantitative estimation of the contribution of scattering is complicated by the complex and 3D structure of the fibrous mat, on the base of the calculated angular distribution of the scattering of individual fibers this is expected to be of the order of few tens of percents, which can be further increased by multiple scattering processes. Such enhancement is compatible with a *weak-scattering regime* [71], due to the low refractive index of the PMMA ($n$=1.49). The use of high refractive index nanofibers [72], which in principle can operate in the *strong-scattering regime*, as substrate for Raman spectroscopy might significantly enhance the contribution of scattering and the increase of Raman signals.

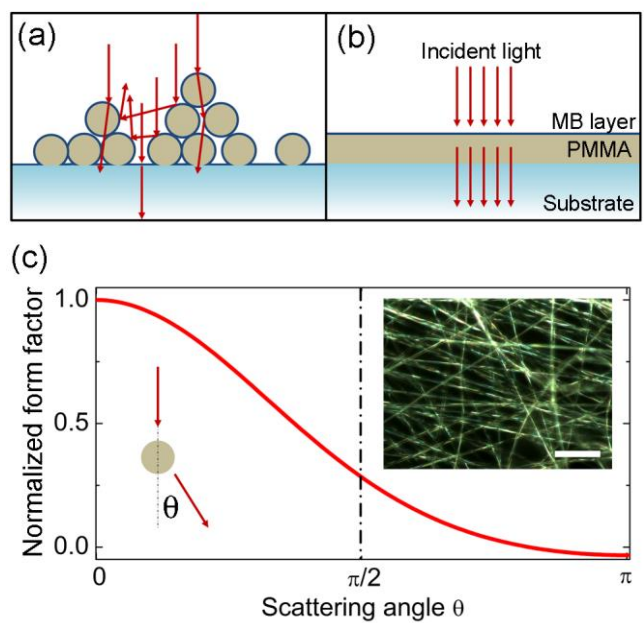

**Figure 3**. (a)-(b) Schematics depicting the interaction of light with nanofibers (a) and with a film (b). The arrows schematize the incident light rays. The direction of propagation of the incident light is almost unchanged in a film, whereas scattering by the fibers deviates the light rays from their original direction, increasing the probability of interaction with the MB layer (blue continuous line). (c) Angular dependence of the normalized scattering form factor at normal incidence with respect to the nanofiber axis, as schematized in the bottom-left scheme. The calculation is performed in the Rayleigh-Gans approximation by assuming a nanofiber with diameter of 400 nm, with refractive index of PMMA ($n$=1.49) surrounded by air. The intensity of light scattered at an angle θ by the nanofiber is proportional to the displayed form factor. Top-right inset: dark field optical image of a mat of PMMA nanofibers. Scale bar: 20 μm. Similar imaging on PMMA spin cast films leads to dark micrographs due to the very low intensity of diffused light.

## SERS spectra on nanofibers

The possibility of further enhancing Raman signals combining the fibrous substrates with metallic nanoparticles is investigated by depositing GNRs on top of either PMMA nanofibers mats or flat substrates without nanofibers, which are here used for control experiments. Figure 4a,b compares the collected SERS spectra of MB molecules at $10^{-4}$ M concentration as adsorbed on GNRs that are drop cast on a mat of nanofibers (Fig. 4a), or on a spin cast PMMA film (Fig. 4b). Taking as a reference the peak at 1618 $cm^{-1}$, for both fibers and films the SERS signal ($I_{SERS}$) is ~ 10 times more intense than the Raman one ($I_{Raman}$) measured on the substrates without GNRs (Fig. 2a and 2b, respectively). In addition, comparing the spectra of Fig. 4a and Fig. 4b, one finds a tenfold amplification of the SERS signal in presence of nanofibers. The SERS spectra of MB molecules on GNR-coated fibers with various diameter (300-800 nm) are shown in Fig. S4. As further control, we compare SERS spectra of MB molecules at $10^{-6}$ M concentration (Fig. 4c). Also in this case, a tenfold improvement of the SERS signal is found on fibers compared to a flat substrate. For sake of comparison, we recall that the Raman signal of the probe molecule collected under same experimental conditions is hardly visible on the flat substrate without GNRs (bottom line in Fig. 4c).

The amplification factor, calculated as $G = I_{SERS}/I_{Raman}$, is typically addressed to as SERS gain [73], and it measures the absolute signal enhancement that a SERS substrate (here, GNRs) can provide for SERS-active molecules (MB). Such a factor is much smaller than the so-called SERS enhancement factor, *EF*, defined as the scattering enhancement provided by each GNR on single molecules. The latter can be calculated upon normalizing the SERS and the Raman signals to the number of probed molecules in each experiment:

$$EF = \frac{I_{SERS}}{I_{Raman}} \frac{n_{Raman}}{n_{SERS}} \qquad (1)$$

where $n_{Raman}$ and $n_{SERS}$ are the number of probed molecules in the Raman and SERS experiments, respectively.

The SERS enhancement factor is of the order of $2\times10^2$ (details reported in ESM), which is interesting for weakly coupled rods excited out of resonance, and in agreement with results on nanorods fabricated by electron beam lithography in the uncoupled regime [74]. From the spectra in Fig. 4 we find that *EF* is similar in nanofibers and in films, consistently with the SERS enhancement by weakly coupled gold nanorods. Finally, Fig. 5 displays SERS spectra collected by MB concentrations down to $10^{-8}$ M. The dependence of the SERS intensity of the peak at 1618 cm$^{-1}$ on the MB concentration is shown in the inset of Fig. 5, highlighting a linear trend. These findings further support the effectiveness of the GNR-coated nanofibers as substrates for SERS applications.

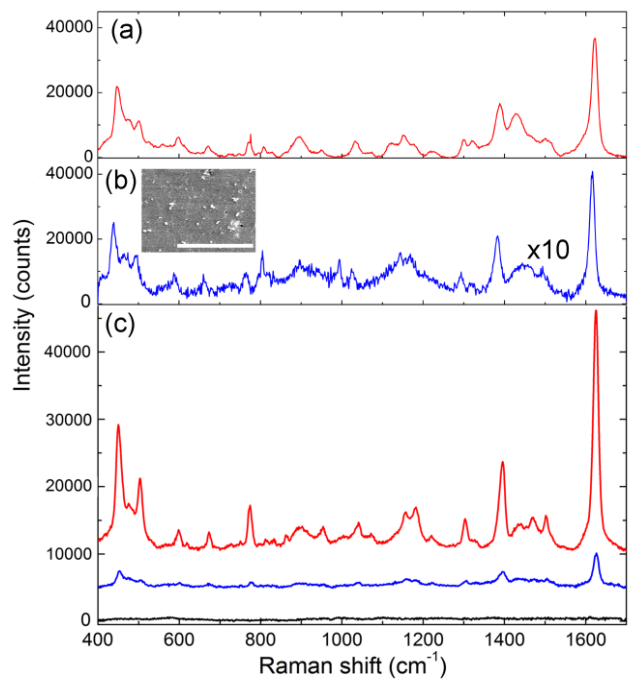

**Figure 4**. (a) SERS spectra of MB molecules ($10^{-4}$ M) adsorbed on a GNR-coated mat of electrospun PMMA nanofibers. As control, spectra of MB molecules ($10^{-4}$ M) adsorbed on GNRs drop cast on a PMMA film are shown in (b). Integration time: 30 s. The spectrum displayed in (b) is multiplied by a factor 10 for better clarity. Inset: SEM micrograph of a PMMA flat film coated with GNRs. Scale bar: 5 μm. (c) Comparison of SERS spectra obtained by MB molecules at $10^{-6}$ M, on a GNR-coated nanofiber mat (top curve) and on GNRs drop cast on a flat, glass substrate (middle curve). Bottom line: Raman signal from MB on flat substrates without GNRs. Integration time: 10 s.

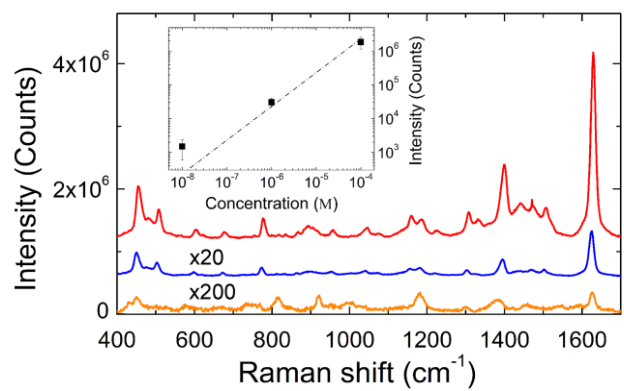

**Figure 5**. Exemplary SERS spectra of MB molecules adsorbed on GNR-coated PMMA nanofiber mats at a concentration of $10^{-4}$ M (top curve), $10^{-6}$ M (middle curve), and $10^{-8}$ M (bottom curve), respectively. The spectra at a concentration of $10^{-6}$ M and $10^{-8}$ M are multiplied by a factor 20 and 200, respectively, for better clarity. Inset: SERS intensity of the 1618 $cm^{-1}$ peak *vs.* MB concentration. Each data point is averaged over ten measurements in different spatial positions on the GNR-coated nanofiber substrate. Error bars represent the standard deviation from the mean values. Dashed line: linear fit to the data.

**Conclusions**

In summary, polymer nanofibers for Raman and SERS applications were produced by electrospinning and the resulting enhancement of the signals from MB cast from aqueous solutions was analyzed. Raman spectra of MB probe molecule adsorbed on nanofiber mats are enhanced by about one order of magnitude compared to films. Geometrical and optical effects contribute to such improvement, the former being related to the increased surface of the complex 3D network formed by the nanofibers, the latter being due to scattering of the incident light from the nanofibers. This phenomenon effectively increases the path of light in the samples, namely the probability of achieving a Raman signal from the probe molecules. Similarly, an enhancement of the intensity of SERS spectra by one order of magnitude is found on nanofibers compared to flat substrates. SERS spectra at a molecular concentration down to $10^{-8}$ M are detected, together with an enhancement factor of the order of $2 \times 10^{2}$ following drop cast deposition of commercial GNRs. In perspective these mechanisms can be further improved, by engineering the nanofiber mats to provide trapping of light inside the fiber network, thus maximizing the effect of multiple scattering.

**Acknowledgments**

We thank A. Foti, B. Fazio, and M. G. Donato for fruitful discussions. We acknowledge the Italian Ministry of Education, University and Research (MIUR), "Programma Operativo Nazionale Ricerca e Competitività" 2007-2013 e "Piano di Azione e Coesione", project PAC02L3_00087 "SOCIAL-NANO". A.C., M.M., L.P. and D.P. also thank the support from the European Research Council under the European Union's Seventh Framework Programme (FP/2007-2013)/ERC Grant Agreement n. 306357 ("NANO-JETS"). The Apulia Networks of Public Research Laboratories Wafitech (09) and M. I. T. T. (13) are also acknowledged.

**Conflict of Interest**: The authors declare no competing financial interest. A. C. and L. P. are the founders of Soft Materials and Technologies SRL.

## ELECTRONIC SUPPLEMENTARY MATERIAL

# Surface-enhanced Raman spectroscopy in 3D electrospun nanofiber mats coated with gold nanorods

Andrea Camposeo,[1,2,*] Donatella Spadaro,[3] Davide Magrì,[2] Maria Moffa,[1] Pietro G. Gucciardi,[3]

Luana Persano,[1,2,*] Onofrio M. Maragò,[3,*] Dario Pisignano[1,4]

[1]*Istituto Nanoscienze-CNR, Euromediterranean Center for Nanomaterial Modelling and Technology (ECMT), via Arnesano, I-73100, Lecce, Italy.*

[2]*Soft Materials and Technologies SRL, via Arnesano 16, I-73100, Lecce, Italy.*

[3]*CNR-IPCF, Istituto per i Processi Chimico-Fisici, I-98158, Messina, Italy*

[4]*Dipartimento di Matematica e Fisica "Ennio De Giorgi", Università del Salento, via Arnesano, I-73100, Lecce, Italy.*

*Corresponding authors: A.C.: andrea.camposeo@nano.cnr.it; O. M. M.: marago@ipcf.cnr.it; L.P.: luana.persano@nano.cnr.it

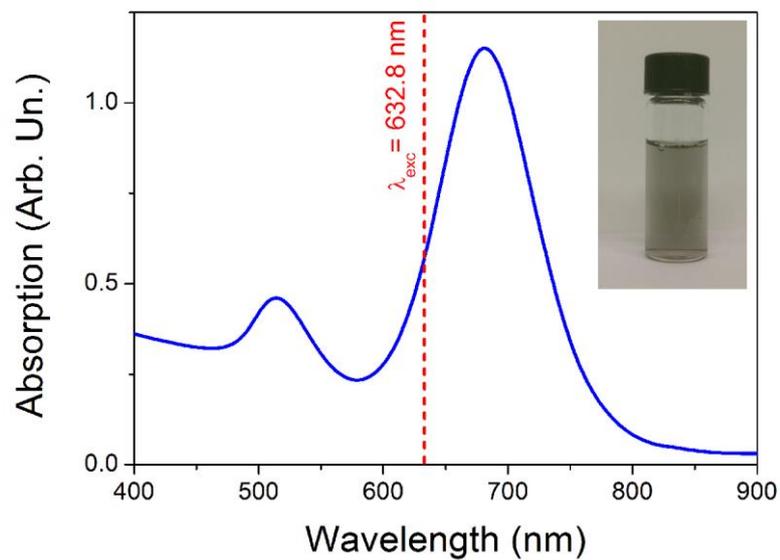

**Figure S1**. UV-visible absorption spectrum of an aqueous dispersion of Au nanorods. The vertical dashed line highlights the wavelength of the excitation laser ($\lambda_{exc}$=632.8 nm) used for Raman and SERS measurements. Inset: photograph of the dispersion.

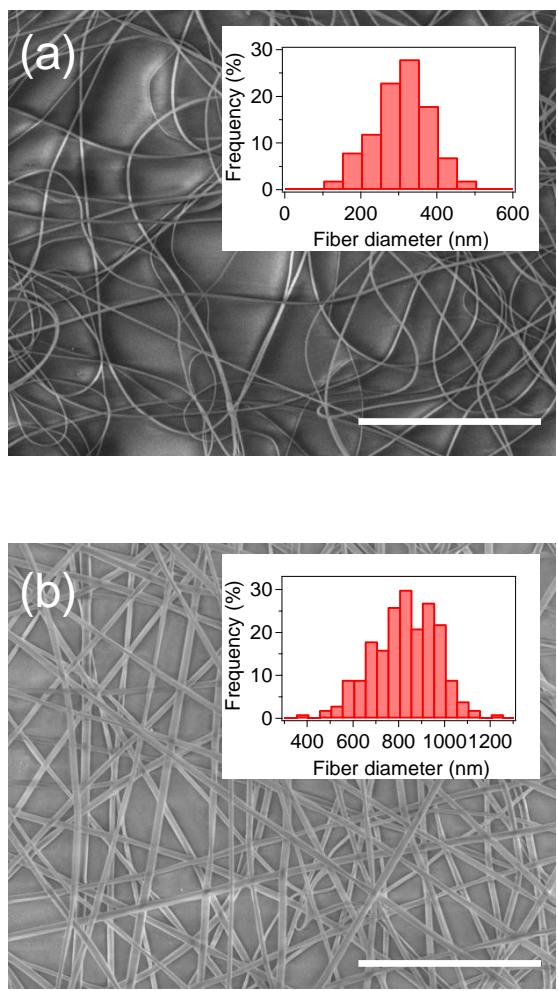

**Figure S2**. (a)-(b) Scanning electron microscopy (SEM) micrographs of PMMA fibers electrospun by solutions with polymer concentration of $2.6 \times 10^{-3}$ M (a) and $2.9 \times 10^{-3}$ M (b), respectively. Scale bars: 20 μm. Fibers shown in panel (a) are obtained by adding tetrabutylammonium iodide to the solution (TBAI:PMMA 1:10 weight:weight relative concentration). Insets: corresponding distributions of fiber diameters. Average diameters = 300 nm (a) and 800 nm (b), respectively.

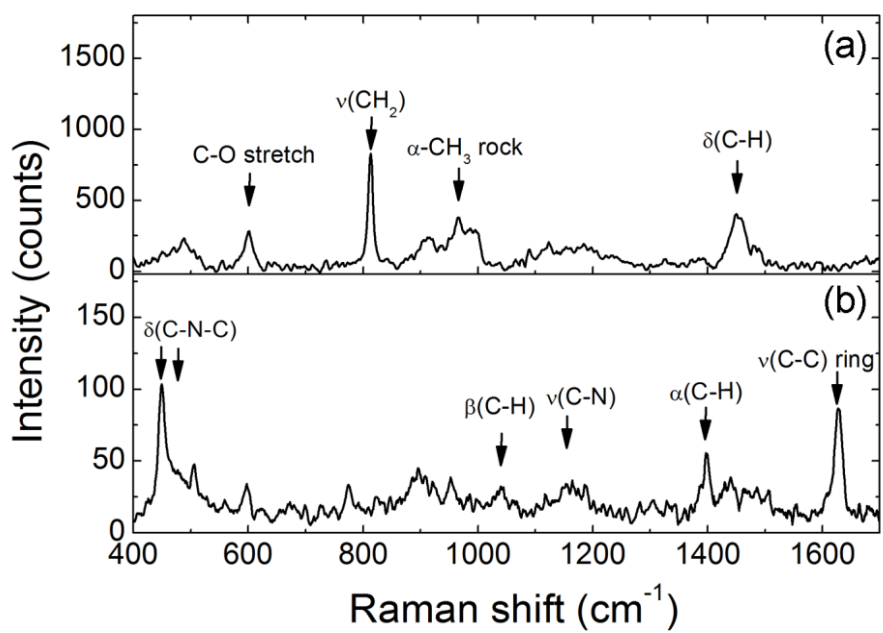

**Figure S3**. Raman spectrum of the PMMA nanofibers (a) and of a drop-cast aqueous solution of methylene blue (MB) on a glass substrate (b). The assignments of the most representative peaks of PMMA and MB are indicated.

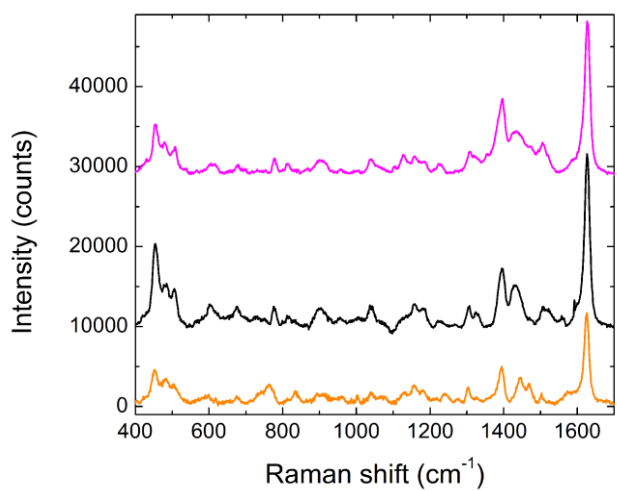

**Figure S4**. SERS spectra of MB molecules on GNR-coated fibers with different average diameters. From bottom to top the average fiber diameter is: 300 nm, 400 nm and 800 nm. The spectra are collected by an integration time of 10 s and are vertically shifted for better clarity.

**SERS enhancement factor calculation**

In order to estimate the SERS enhancement factor ($EF$, Eq. 1), the number of probed molecules ($n_{Raman}$ and $n_{SERS}$) has to be determined. An estimate can be obtained assuming that all the molecules within the laser spot area do contribute to the Raman signal ($I_{Raman}$). Conversely, only the molecules adsorbed on the gold nanorods surface area $A_{GNR}$ will contribute to SERS signal ($I_{SERS}$). Here $A_{GNR} = \bar{n}_{GNR} \, \pi \, d_{GNR} \, l_{GNR}$, where $\bar{n}_{GNR}$ is the average number of GNRs illuminated by the laser spot, $d_{GNR}$ and $l_{GNR}$ are the GNRs diameter and length, respectively. Without loss of generality we can calculate the $EF$ considering the film geometry, for which the laser spot area is given by: $A_{las} = \pi \, d_{las}^{2}/4$, where $d_{las} \sim 1.5$ µm is the laser spot diameter. We assume that the surface density of MB molecules ($\sigma$, number of absorbed molecules per unit area) is the same for the PMMA film and the GNRs. Under this approximation the number of probed molecules in Raman and SERS experiments is given by: $n_{Raman,SERS} = \sigma \, A_{Las,GNR}$, which allows us to estimate their ratio as:

$$\frac{n_{Raman}}{n_{SERS}} \sim \frac{d_{las}^{2}}{4 \, \bar{n}_{GNR} \, d_{GNR} \, l_{GNR}} \sim 300/\bar{n}_{GNR} \qquad (S1)$$

The average number of GNRs, $\bar{n}_{GNR}$, illuminated in the SERS experiments can be estimated from the SEM micrographs of PMMA films coated with GNRs (inset of Figure 4b), which provides an average value of 15 rods/µm². By using such value of GNRs density one finds an enhancement factor $EF \sim 2\times10^{2}$.